\documentclass{PoS}

\newcommand{\diag}{{\rm \mbox{{  diag}}}}
\newcommand{\ini}{{\rm \mbox{{\scriptsize  ini}}}}

\title{Out-of-equilibrium energy flow and steady state configurations in AdS/CFT}

\ShortTitle{Out-of-equilibrium energy flow and steady state configurations in AdS/CFT}

\author{\speaker{Eugenio Meg\'{\i}as} \\
        Max-Planck-Institut f\"ur Physik (Werner-Heisenberg-Institut), \\
        F\"ohringer Ring 6, D-80805, Munich, Germany \\
        E-mail: \email{emegias@mppmu.mpg.de}
}

\abstract{We study out-of-equilibrium energy flow in a strongly coupled system by using the AdS/CFT correspondence. In particular, we describe the appearance of a steady state connecting two asymptotic equilibrium systems. We obtain results within the linear response regime.}

\FullConference{The European Physical Society Conference on High Energy Physics\\
		22--29 July 2015\\
		Vienna, Austria}

\begin{document}

\section{Introduction}
\label{sec:Introduction}

Hydrodynamics has been extensively used to study systems which are close to equilibrium. It is based on the assumption that the mean free path (time) of particles is much shorter than the characteristic size (time scale) of the system, and the result can be organized in a gradient expansion, also called hydrodynamic expansion~\cite{Kovtun:2012rj}. However, this approach has some limitations, as it usually fails to describe far from equilibrium systems appearing in many branches in physics: some situations are the initial stages of the Quark-Gluon plasma thermalization~\cite{Ishii:2015gia}, quenches in some condensed matter systems and fluctuations in the fractional Hall effect~\cite{Polkovnikov:2010yn}. Basic approaches to these systems include the computation of real time dynamics directly, and it is generally expected that they evolve towards a hydrodynamic regime at late times. However understanding far-from-equilibrium physics is a notoriously challenging problem, so that we have to resort to much simpler systems to have a better understanding of these phenomena. An interesting yet potentially tractable class of non-equilibrium configurations are the steady state flows: examples are the electric current in a conductor driven by an external electric field, or the heat current driven by a temperature gradient~\cite{Bernard:2012je,Chang:2013gba,Bhaseen:2013ypa}. These are time-independent configurations but they do not correspond to equilibrium.

These studies are particularly interesting in strongly coupled systems. The AdS/CFT correspondence constitutes a powerful tool that helped to establish some universal properties of the hydrodynamics of quantum systems, the most famous one being the celebrated lower bound for the shear viscosity to entropy density ratio, $\hbar/4\pi$~\cite{Policastro:2001yc}. A strong motivation to apply the AdS/CFT correspondence to far-from-equilibrium dynamics is that it might establish as well some universal properties of these systems, in particular it could give some insight about the organizing principles out-of-equilibrium and the possible existence and characterization of universality classes. In this work we will study, within the AdS/CFT correspondence, a particular out-of-equilibrium steady state configuration consisting of a heat current between two asymptotic equilibrium systems, its formation and time evolution.

\section{A universal regime of thermal transport: steady state formation}
\label{sec:universal}

It was shown in Ref.~\cite{Bernard:2012je} that in a class of Conformal Field Theories (CFT) in $(1+1)$-dim a homogeneous steady state exists, and a universal formula for the heat flow was derived. These results were generalized later to higher dimensions in Refs.~\cite{Chang:2013gba,Bhaseen:2013ypa}. We will present in this section some of the properties of these configurations.

\begin{figure*}[htb]
\begin{tabular}{ccc}
\includegraphics[width=63mm]{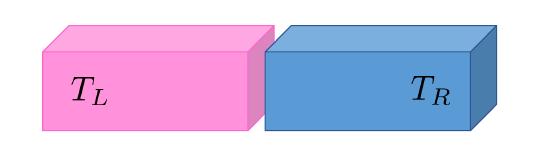} &
\hspace{-0.5cm}\includegraphics[width=20mm]{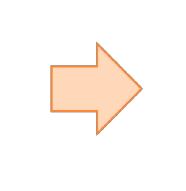} &
\hspace{-0.3cm}\includegraphics[width=63mm]{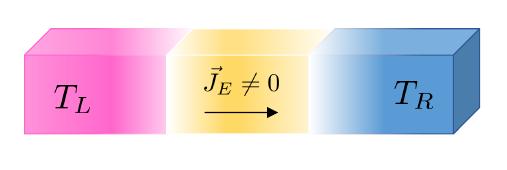} \\
\end{tabular}
\caption{Two isolated systems initially at equilibrium are put in contact at $t=0$. A spatially homogeneous non-equilibrium steady state develops at late times, and it carries an energy current $J_E = \langle T^{tx}\rangle_s$.}
\label{fig:system}
\end{figure*}

Let us consider two thermal reservoirs in $d$-dim, each of them initially at equilibrium but at different temperatures, $T_L$ and $T_R$. These systems are put in contact at the initial time $t=0$. Such a physical situation is presented in Fig.~\ref{fig:system}. 
The initial energy density reads
\begin{equation}
\varepsilon(x,t=0) = (d-1) a_d \left[ T_L^{d} \Theta(-x) + T_R^{d} \Theta(x) \right] \,,
\end{equation}
where $a_d$ depends on the number of degrees of freedom in the CFT. After bringing the two systems into thermal contact, a spatially homogeneous steady state develops near the interface, carrying a heat flow $J_E$ which transfers energy from the hottest to the coldest system. As discussed in Ref.~\cite{Bhaseen:2013ypa}, the steady state configuration in the CFT can be described by the Lorentz boosted stress tensor
\begin{equation}
\langle T^{\mu\nu} \rangle_s = a_d T^{d} \left( \eta^{\mu\nu} + d u^\mu u^\nu \right) \,, \label{eq:Tmunu}
\end{equation}
where $\eta^{\mu\nu} = \diag(-1,1,\cdots,1)$ and the fluid velocity is $u^\mu = (\cosh\theta, \sinh\theta,0, \cdots,0)$. An intuitive picture is that the boost leads to a Doppler shift of the thermal radiation between the left and right-movers, i.e.
\begin{equation}
T_L = T_s \, e^{\theta} \,, \qquad T_R = T_s \, e^{-\theta}  \,, \qquad T_s = \sqrt{T_L T_R} \,, \label{eq:TLTRTs}
\end{equation}
where $T_s$ is the temperature of the steady state. By using the conservation of energy and momentum and traceless of the stress tensor in the CFT,
\begin{equation}
\partial_\mu \langle T^{\mu\nu}\rangle =0  \,, \qquad \langle T^{\mu}_{\mu}\rangle = 0  \,, \label{eq:hydro_eom}
\end{equation}
it has been obtained solutions in a perfect fluid consisting of ``{\it shockwaves}'' emanating from the interface~\cite{Smoller:1993}. Enforcing Eq.~(\ref{eq:hydro_eom}) across the shocks one gets the following result for the energy current in the steady state~\cite{Chang:2013gba,Bhaseen:2013ypa}
\begin{equation}
\langle T^{tx} \rangle_s = \frac{1}{2} d a_d T^d \sinh (2\theta) =  a_d \left( \frac{T_L^{d} - T_R^{d}}{u_L + u_R} \right) \,.
\end{equation}
In the rest of the manuscript we will study the time evolution of this system by using a simple holographic model. Our goal is to describe, not only the steady state, but all the space-time regime.

\section{Energy flow and time evolution of steady states in AdS/CFT}
\label{sec:AdS_CFT}

In this section we will present the simplest holographic model to study the system described above, and find a solution by linearizing the problem. Other out-of-equilibrium stationary configurations in strongly coupled systems can be found in e.g.~\cite{Fischetti:2012vt,Emparan:2013fha} and references therein.

\subsection{Holographic model}
\label{subsec:holographic_model}

 Let us consider the Einstein-Hilbert action in $(d+1)$-dim given by 
\begin{equation}
S= {1 \over 16\pi G} \int d^{d+1}x \sqrt{-g}\left\{  R - 2\Lambda \right\}  \,,
\label{eq:action}
\end{equation}
where~$\Lambda = -d(d-1)/2$ is a negative cosmological constant. The equations of motion write
\begin{equation}
R_{\mu\nu} - \frac{1}{2}g_{\mu\nu} R + g_{\mu\nu} \Lambda = 0 \,, \qquad \mu, \nu = 1,\cdots, d+1 \,. \label{eq:eom}
\end{equation}
For the moment we will restrict to $d=3$ and choose $x^\mu = (t,x,y,z)$, where $z$ is the holographic coordinate. The kind of solution we are looking for is a boosted black hole with metric
\begin{equation}
 \hat g_{\mu\nu} = \frac{1}{z^2}
\left(
\begin{array}{cccc}
 \left( \sinh[\theta]^2 - \cosh[\theta]^2 f \right) &  \frac{1}{2}\sinh[2\theta] (f-1)  & 0 & 0  \\
  \frac{1}{2} \sinh[2\theta] (f-1)  & \left( \cosh[\theta]^2 - \sinh[\theta]^2 f \right) & 0 & 0  \\
 0 & 0 & 1 & 0  \\
 0 & 0 & 0 & \frac{1}{f}
\end{array}
\right) \,, \label{eq:g}
\end{equation}
where $f = 1-\left(\frac{z}{z_h}\right)^d$. This metric is a solution of the equations of motion~(\ref{eq:eom}) as long as $z_h$ and $\theta$ are constants in space-time. In this case $z_h = d/(4\pi T)$ where $T$ is the unboosted temperature of the black hole. However, our goal is to promote these parameters to be space-time dependent, i.e. $z_h = z_h(t,x)$ and $\theta = \theta(t,x)$, and for that one needs to include some extra degrees of freedom.

\subsection{Linearization}
\label{subsec:linearization}

A convenient way to proceed is by linearizing the problem. The idea is the following: let us consider that the space-time dependence in $z_h$ and $\theta$ enters in the form
\begin{equation}
z_h(t,x) = z_{h(0)} + \epsilon z_{h(1)}(t,x) + \cdots  \,, \qquad \theta(t,x) = \theta_{(0)} + \epsilon \theta_{(1)}(t,x) + \cdots  \,. \label{eq:zh_theta}
\end{equation}
This means that we are searching for solutions as an expansion in powers of a formal parameter~$\epsilon$, so that all the space-time dependence is treated as a perturbation around the background $z_{h(0)}$ and~$\theta_{(0)}$, which we keep constant and uniform. One can always replace $\epsilon \to 1$ at the end. As mentioned above, it is necessary to add extra contributions to the metric to compensate the effects of $z_{h(1)}(t,x)$ and $\theta_{(1)}(t,x)$ in the equations of motion. In particular, we consider a metric of the form
\begin{equation}
 g_{\mu\nu}(t,x,y,z) =  \hat g_{\mu\nu}(t,x,y,z)
+
\epsilon \left(
\begin{array}{cccc}
  \delta g_{tt}(t,x,z) &  \delta g_{tx}(t,x,z) & 0 & 0  \\
   \delta g_{tx}(t,x,z)  &  \delta g_{xx}(t,x,z) & 0 & 0  \\
 0 & 0 & 0 & 0  \\
 0 & 0 & 0 &  \delta g_{zz}(t,x,z)
\end{array}
\right) \,, \label{eq:gdg}
\end{equation} 
where $\hat g_{\mu\nu}(t,x,y,z)$ is given by Eq.~(\ref{eq:g}) with the space-time dependent parameters of Eq.~(\ref{eq:zh_theta}). A~solution would be to set $\delta g_{\mu\nu}$ as the value which exactly cancels the order ${\cal O}(\epsilon)$ in an expansion of~$\hat g_{\mu\nu}$. However, this leads to the trivial solution corresponding to a uniform system, valid only to describe the steady state regime. The non-trivial solution we find can be written as a near boundary expansion, and the result up to the first non-vanishing order reads
\begin{eqnarray}
\delta g_{tt}(t,x,z) &=&\frac{1}{2z_{h(0)}^4}(\partial_t^2 z_{h(1)}) z^3 + {\cal O}(z^5) =  \delta g_{zz}(t,x,z)  \,, \\
\delta g_{tx}(t,x,z) &=& \frac{3}{10z_{h(0)}^4}(\partial_t\partial_x z_{h(1)}) z^3 + {\cal O}(z^5) \,, \\
\delta g_{xx}(t,x,z) &=& \frac{2}{5z_{h(0)}^4}(\partial_t^2 z_{h(1)}) z^3 + {\cal O}(z^5) \,.
\end{eqnarray}
Using this metric, the relation between temperature $T$ and horizon $z_h$ is~\footnote{This follows from a {\it non-trivial} computation by using the killing vector of the metric~(\ref{eq:gdg}). Note that, at least up to order ${\cal O}(\epsilon)$, this expression is equivalent to the more familiar form $T(t,x) =  \frac{d}{4\pi z_h(t,x) }$, where $z_h(t,x)$ is given by Eq.~(\ref{eq:zh_theta}).}
\begin{equation}
T(t,x) = \frac{d}{4\pi z_{h(0)}}\left(1 - \epsilon \frac{z_{h(1)}(t,x)}{z_{h(0)}} + {\cal O}(\epsilon^2) \right) \,. \label{eq:Temptx}
\end{equation}
Finally, the Einstein equations of motion reduce to the following two equations
\begin{eqnarray}
0 &=& \partial_t^2 z_{h(1)}(t,x) -  c_s^2 \partial_x^2 z_{h(1)}(t,x) \,, \label{eq:R22} \\
0 &=& \partial_t z_{h(1)}(t,x)  - c_s^2 z_{h(0)} \partial_x \theta_{(1)}(t,x) \,, \label{eq:R44}
\end{eqnarray}
where $c_s^2 = 1/2$.  To arrive at this simple result we have set $\theta_{(0)}=0$.~\footnote{Note that in a uniform system one expects a vanishing value of the energy flow, and this can only be achieved by setting $\theta_{(0)}=0$.} We have made this analysis for other space-time dimensions, and in every case we get Eqs.~(\ref{eq:R22})-(\ref{eq:R44}) with $c_s^2 = 1/(d-1)$.

\subsection{Solution of the equations of motion}
\label{subsec:solution_eom}

\begin{figure*}[htb]
\includegraphics[width=70mm]{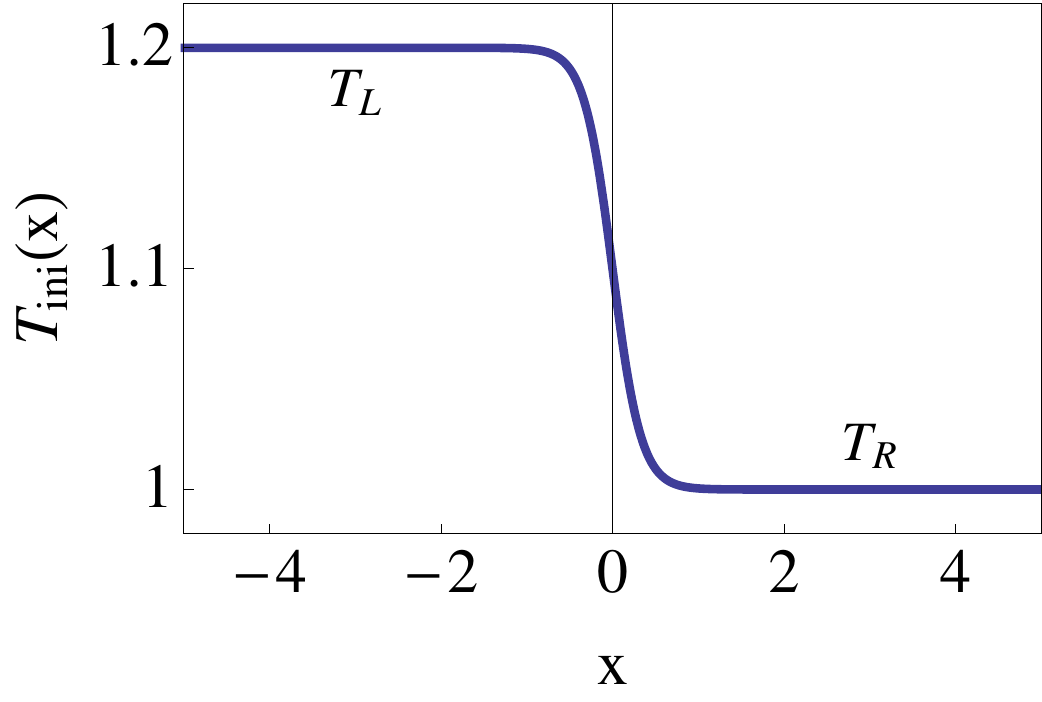} \hspace{2pc}%
\begin{minipage}{15pc}
\vspace{-3cm}\caption{Initial profile given by Eq.~(\protect\ref{eq:Tini2}) with $T_L = 1.2$, $T_R = 1$ and $\alpha = 3$.}
\label{fig:Tini}
\end{minipage}
\end{figure*}

\begin{figure*}[htb]
\begin{tabular}{cc}
\includegraphics[width=70mm]{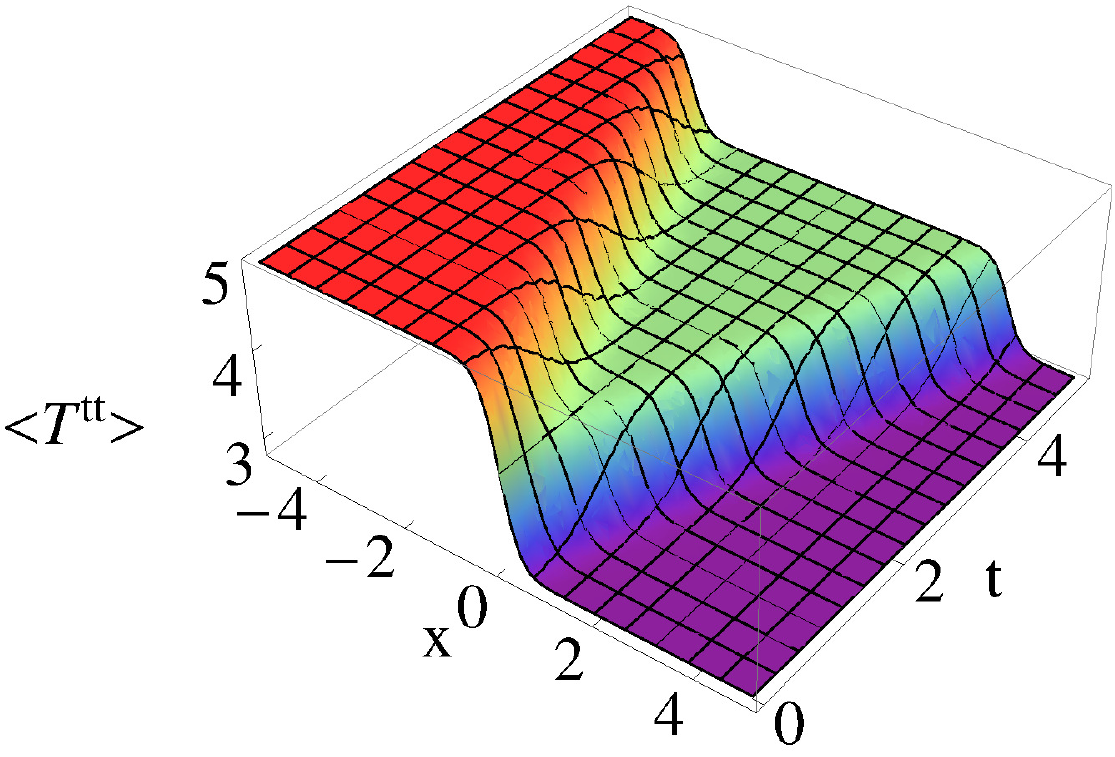} &
\includegraphics[width=70mm]{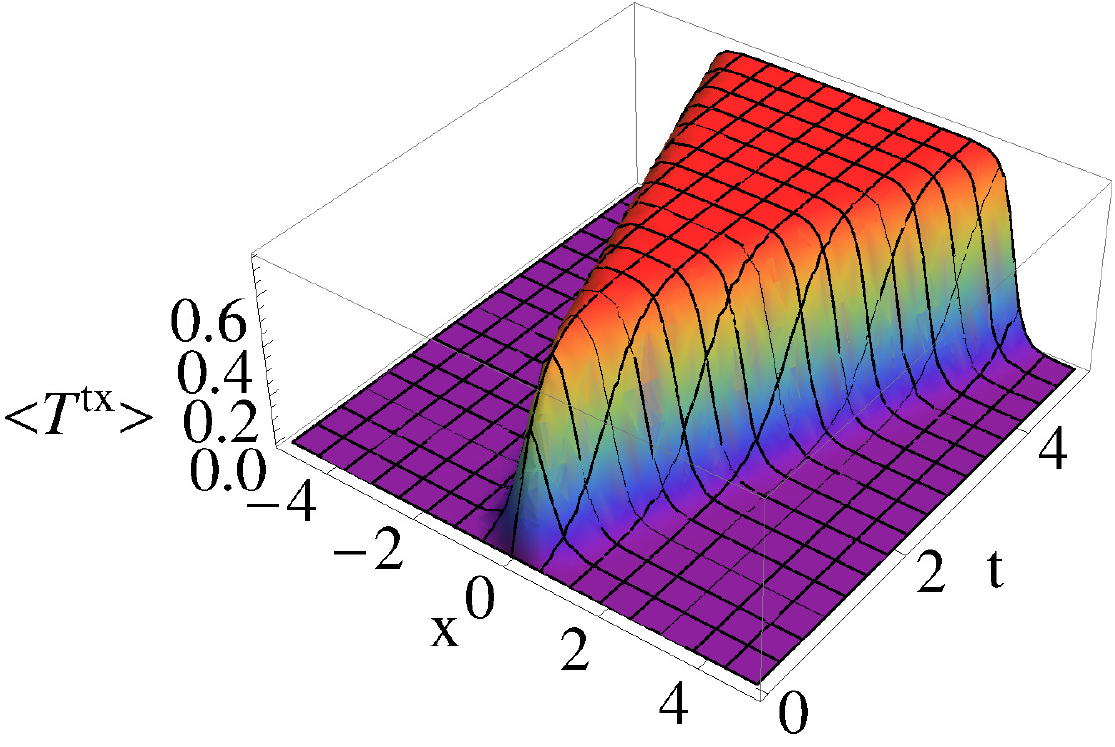} \\
\end{tabular}
\caption{Energy density and energy current computed with the linearized solution given by Eqs.~(\protect\ref{eq:Ttt}) and (\protect\ref{eq:Ttx}) respectively. It is used the initial profile of Fig.~\protect\ref{fig:Tini}. We have considered $d=3$ and set $G=1$.}
\label{fig:TttTtx}
\end{figure*}

The general solution of the equations of motion, Eqs.~(\ref{eq:R22})-(\ref{eq:R44}), can be written as
\begin{eqnarray}
z_{h(1)}(t,x) &=& F_1\left(x + c_s t\right) + F_2\left(x - c_s t\right)  \,, \label{eq:zh1} \\
\theta_{(1)}(t,x) &=&   \frac{1}{c_s z_{h(0)}} \left[ F_1\left(x + c_s t \right) - F_2\left(x - c_s t \right) \right]  \,, \label{eq:theta1}
\end{eqnarray}
where $F_1(v)$ and $F_2(v)$ are arbitrary functions of their arguments. One can obtain explicit values for these functions after imposing the appropriate boundary conditions. In particular, if one assumes some initial profile for the temperature, i.e. $T_{\ini}(x) \equiv T(t=0,x)$, then from Eqs.~(\ref{eq:Temptx}) and (\ref{eq:zh1}) one gets
\begin{equation}
T_{\ini}(v) = \frac{d}{4\pi z_{h(0)}} \left[ 1 - \frac{\epsilon}{z_{h(0)}} (F_1(v) + F_2(v) ) \right] \,. \label{eq:Tini}
\end{equation}
Following the rules of the AdS/CFT correspondence~\cite{deHaro:2000xn}, we can compute all the components of the energy-momentum tensor~$\langle T^{\mu\nu} \rangle$ in the CFT from the dual gravity metric of Sec.~\ref{subsec:linearization}. The other boundary condition then follows from the requirement that there is no energy flow at $t=0$, i.e. $\langle T^{tx}(t=0,x)\rangle = 0$ for $x \in (-\infty, +\infty)$, and this leads to $F_1(v) = F_2(v)$. Finally one gets
\begin{eqnarray}
\langle T^{tt}(t,x) \rangle &=& \frac{(d-1)}{8 G} \frac{1}{z_{h(0)}^{d-1}} \left[ -\frac{(d-1)}{2 \pi z_{h(0)}}  + T_\ini(x+ c_s t) + T_\ini(x - c_s t)  \right]   \,,  \label{eq:Ttt} \\
\langle T^{tx}(t,x) \rangle &=& - \frac{1}{c_s} \frac{1}{8 G } \frac{1}{z_{h(0)}^{d-1}} \left[ T_\ini(x+ c_s t) - T_\ini(x - c_s t) \right]   \,, \label{eq:Ttx}
\end{eqnarray}
for the energy density and energy flow, respectively. This solution leads to the existence of ``{\it shockwaves}'' propagating at speed $c_s$, and it fulfills Eq.~(\ref{eq:hydro_eom}).~\footnote{See Refs.~\cite{Smoller:1993,Chang:2013gba} for an alternative derivation of this solution by using hydrodynamic considerations.} For numerical computations we choose the following initial profile
\begin{equation}
T_\ini(x) = \frac{T_R + T_L}{2} + \frac{T_R-T_L}{2}\tanh(\alpha x) \,, \label{eq:Tini2}
\end{equation}
which tends to the stepwise function $T_\ini(x) \to T_L \Theta(-x) + T_R \Theta(x)$ in the limit $\alpha \to \infty$. If one makes the identification $z_{h(0)} = \frac{d}{2\pi (T_L + T_R)}$, it is possible to see that corrections of order ${\cal O}(\epsilon)$ are always proportional to factors $\propto \frac{T_R - T_L}{T_R + T_L}$. This illustrates the fact that the $\epsilon$-expansion of Sec.~\ref{subsec:linearization} is equivalent to a small gradient expansion, i.e. $\Big|\frac{T_R - T_L}{T_R + T_L}\Big| \ll 1$, and ultimately to linearized hydrodynamics. We find that the steady state temperature is $T_s \equiv T(t \to +\infty , x) = \frac{T_R + T_L}{2} + {\cal O}(\epsilon^2)$, which is none other than the expansion of Eq.~(\ref{eq:TLTRTs}) in powers of $\epsilon$ with $T_L = T_R \cdot (1+\epsilon)$. We display in Figs.~\ref{fig:Tini} and \ref{fig:TttTtx} the numerical result with these formulas. The formation of the steady state and the propagation of the shockwaves is  properly described in the regime of small difference of temperatures.

\section{Conclusions and discussion}
\label{sec:conclusions}

We have studied a holographic model for out-of-equilibrium energy flow that allows us to describe the energy transport in a system consisting of two thermal reservoirs initially isolated. The computation is performed by finding a boosted black hole solution of the equations of motion, with space-time dependent contributions. This leads to a description of the formation of a steady state and the propagation of shockwaves, already anticipated in~\cite{Smoller:1993,Chang:2013gba,Bhaseen:2013ypa}. Our main assumption has been a linearization of the theory which turns out to be equivalent to a small gradient expansion. 

There remain some open questions. It would be interesting to perform an analysis beyond the linear response regime, i.e. for $0 < T_R/T_L < 1$. This demands a full numerical solution of the equations of motion, see e.g.~\cite{Chesler:2013lia}. A recent work which studies the time evolution of asymptotically AdS black branes in this line is~\cite{Amado:2015uza}. Another target is the existence of other possible solutions: it is argued in~\cite{Chang:2013gba} that other branch of solutions appears in addition to the thermodynamic branch discussed in this work, although it is still not clear its physical role. Finally, it deserves to be pursued a study of information flow, i.e. how the information gets exchanged between two systems which are initially isolated. A convenient quantity for this is the entanglement entropy, defined holographically as the area of a minimal surface extending from some predefined surface on the boundary into the bulk, and which is a generalization of the Bekenstein-Hawking entropy formula~\cite{Ryu:2006ef,Hubeny:2007xt}. Some works of entanglement entropy in time-dependent systems related to the one studied here are~\cite{AbajoArrastia:2010yt,Ecker:2015kna}. These and other issues will be addressed in a forthcoming publication~\cite{Megias:2015inprogress}.

\acknowledgments 


I would like to thank J. Erdmenger, D. Fern\'andez, M. Flory and A.K. Straub for collaboration on related topics, and M. Ammon, C. Ecker, D. Grumiller, E. Kiritsis, E. L\'opez and A. Yarom for enlightening discussions. I thank the String Theory Group at the Technion Israel Institute of Technology for hospitality during the process of writing this manuscript, and the German-Israeli Foundation grant GIF-1156 for travel support. Research  supported by the European Union under a Marie Curie Intra-European Fellowship (FP7-PEOPLE-2013-IEF) project PIEF-GA-2013-623006.


\end{document}